\documentclass[aps,preprint]{revtex4}%
\usepackage{amsfonts}
\usepackage{amsmath}
\usepackage{amssymb}
\usepackage{graphicx}%
\begin{document}
\title{How kinetics drives the two- to three-dimensional transition in semiconductor
strained heterostructures: the case of InAs/GaAs(001).}
\author{F. Arciprete\thanks{Corresponding author, E-mail:
fabrizio.arciprete@roma2.infn.it}, E. Placidi, V. Sessi, M. Fanfoni, F.
Patella and A. Balzarotti}
\affiliation{Dipartimento di Fisica, Universit\`{a} di Roma \textquotedblleft Tor
Vergata\textquotedblright, Via della Ricerca Scientifica 1, I-00133 Roma, Italy.}

\begin{abstract}
The two- to three-dimensional growth transition in the InAs/GaAs(001)
heterostructure has been investigated by atomic force microscopy. The kinetics
of the density of three dimensional quantum dots evidences two transition
thresholds at $1.45$ and $1.59$ ML of InAs coverage, corresponding to two
separate families, small and large. Based on the scaling analysis, such
families are characterized by different mechanisms of aggregation, involving
the change of the critical nucleus size. Remarkably, the small ones give rise
to a wealth of "monomers" through the erosion of the step edges, favoring the
explosive nucleation of the large ones. \ \ 

\end{abstract}
\date{}
\maketitle
\preprint{ }
\volumeyear{year}
\volumenumber{number}
\issuenumber{number}
\eid{identifier}
\received[Received text]{date}

\revised[Revised text]{date}

\accepted[Accepted text]{date}

\published[Published text]{date}

\startpage{1}
\endpage{ }

In the previous decade the Stranski-Krastanov (SK) growth mode has been the
subject of many experimental and theoretical studies. In the case of
semiconductor heterostructures which are characterized by a large lattice
mismatch, the strain built up during the layer by layer growth may bring about
a drastic and sudden change in the growth mode; atoms start self-organizing in
three-dimensional (3D) islands called quantum dots
(QDs).\cite{Stangl2004,Joyce2004} InAs/GaAs(001) and Ge/Si(001)
heterostructures are paradigmatic examples of these interfaces. The former
with its lattice mismatch as large as $7\%$, undergoes the 2D-3D transition
for a deposition of InAs lower than $2$ ML.

Many attempts have been made to clarify the mechanisms underlying the 2D-3D
transition. A recent significant result is due to Cullis \textit{et
al.}\cite{Cullis2001,Cullis2002}. By studying the crucial role played by the
In segregation in the 2D-3D transition for the InGaAs/GaAs(001) growth, they
have reached an important conclusion: \textit{When the average In
concentration in the growing layer reaches a value in the range }%
$80-85\%$\textit{, islanding begins}. This conclusion establishes a kind of
thermodynamic constraint in order for the transition to take place. The model
applicability has also been explored in the Ge/Si
heterostructure\cite{Cullis2005,Tersoff2004}. Among other established
facts,\cite{Joyce2004} it is widely accepted, where islanding is concerned,
that the InAs/GaAs(001) film's morphology during the transition is
characterized by two distinct families of QDs: the so called quasi-3D QDs,
whose height and lateral dimensions range from $0.3$ to $0.7$ $%
\operatorname{nm}%
$ and from $10$ to $20$ $%
\operatorname{nm}%
$ respectively, and the mature QDs, whose height is higher than $2$ $%
\operatorname{nm}%
$ and whose lateral dimensions range between $20$ and $30$ $%
\operatorname{nm}%
$. Henceforth the former will be referred to as small and the latter as large
QDs. Small QDs disappear very soon and the surface, before coalescence, turns
out to be dotted with large QDs. Hitherto, what is the role played by the
small QDs has not been well understood, and several conjectures have been put
forward, often on the basis of untargeted data. In this Letter we present an
accurate investigation of the kinetics of both small and large QDs for the
InAs/GaAs(001) heterostructure grown by means of molecular beam epitaxy (MBE).
In particular, we show that small and large QDs give rise to two distinct
families: Remarkably, the small ones do not account for the number of all the
large ones, although, in a certain way, the small QDs favor the explosive
nucleation of the large ones. Hence, although energetics fixes the transition
threshold, kinetics plays a fundamental role in determining its subsequent
evolution. Although related to a specific system, the relevance of this
result, is that it could be common to all highly-mismatched semiconductor
heterostructures characterized by a similar phenomenology.

Because the system evolves within a very narrow range of coverage around the
critical thickness,\cite{Petroff1994,Rama1996,Rama1997,Jones2002,Patella2004}
we took advantage of the intrinsic non-uniformity of the molecular beams to
grow a single sample where the InAs film thickness varies continuously along
the sample surface. In this way a snapshot of the QD evolution is available
and the system can be studied for InAs coverage increments as low as $0.01$
ML. Furthermore, any problem of reproducible growth conditions is overcome.
The investigated sample has been grown by means of an MBE reactor equipped
with Reflection High-Energy Electron Diffraction (RHEED) for monitoring the
growth. Prior to InAs deposition, a GaAs regrowth of approximately $500$ $%
\operatorname{nm}%
$ was performed on the $(001)$ oriented substrate, in As$_{4}$ overflow, at
$590$ $%
\operatorname{{}^{\circ}{\rm C}}%
$ at a rate of $1%
\operatorname{\mu m}%
$/$%
\operatorname{h}%
$. After $10$ $%
\operatorname{min}%
$ post-growth annealing, the temperature was lowered to $500$ $%
\operatorname{{}^{\circ}{\rm C}}%
$ for the InAs deposition. The growth was carried out without rotating the
sample, so as to obtain the afore-mentioned non-homogeneous $2$-inch sample.
The impinging flux increases linearly along the $[110]$ direction of the
substrate,\cite{nota1} from $0.011$ ML/$%
\operatorname{s}%
$ to $0.030$ ML/$%
\operatorname{s}%
$, resulting in InAs coverage ranging from $0.87$ ML to $2.40$ ML for $80$ $%
\operatorname{s}%
$ of growth. The In delivery was cycled in $5$ $%
\operatorname{s}%
$ of evaporation followed by $25$ $%
\operatorname{s}%
$ of growth interruption until the beginning of the 2D-3D transition was
observed by RHEED at the center of the sample.\cite{Patella2004} Atomic force
microscopy (VEECO Multiprobe) was performed in air in the tapping mode by
using non-conductive Si tips, on $20$ different points of the sample for InAs
coverage ranging from $0.87$ to $2.22$ ML.

Fig.1 shows AFM topographies ($1.0$ $%
\operatorname{\mu m}%
$ $\times$ $0.5$ $%
\operatorname{\mu m}%
$) for three significant InAs coverages: $1.54$ ML, $1.57$ and $1.64$ ML in
Fig. 1(a), (b), and (c), respectively. The images reveal a complex morphology
of the WL, \textit{i.e.} 2D islands $1$ ML high, and large terraces one step
high. The first small QDs (Fig. 1(a) and (b)) are recognizable for coverage as
high as $1.45$ ML, whereas at higher InAs deposits the emergence and
subsequent increase of the number of large QDs can be seen (Fig. 1(b) and
(c)). Small QDs nucleate preferentially at the upper-step edges of 2D islands
and terraces (Fig. 1 (a) and (b)) by reason of a favorable strain condition at
those sites. They have been reported several
times\cite{Rama1997,Dasilva2002,Jones2002,Patella2003,Costantini2005} and
often been indicated as simple precursors of large QDs. We have already
pointed out\cite{Patella2003} that this simple picture is unrealistic and we
will show below that the process involves a more complex kinetic mechanism.

The number density evolution of both small and large QDs\cite{nota2} is
summarized in Fig. 1(d) as a function on InAs deposition. The number of the
small QDs begins to increase at $1.45$ ML of InAs deposit and maximizes at
$1.57$ ML reaching the value of $1.1\times10^{10}$ $%
\operatorname{cm}%
^{-2}$. Starting from $1.52$ ML, the number of the large QDs increases
gradually then, between $1.57$ and $1.61$ ML, undergoes a sudden rise,
changing value by an order of magnitude. At higher coverages the density rise
is much slower. The steady and gentle increase observable for coverages higher
than $1.8$ ML is due to the dependence of the density saturation value on
growth rate.\cite{Venables1973,Tomellini1998,Jones2002b} In this region the
density is of the order of $6\times10^{10}$ $%
\operatorname{cm}%
^{-2}$.

The key point lies in understanding the transition process between $1.57$ and
$1.61$ ML. The comparison of the number density of the two QD families rules
out the possibility that the large QDs are merely the evolution of the small
ones, \textit{i.e.} the low density of the small QDs cannot account for the
density evolution of the large ones. To gain an insight into the nature of
these two families we have analyzed the scaling behavior of their size
distribution. As a matter of fact, dynamic scaling theory makes it possible to
determine one of the most significant parameters of film formation governed by
nucleation and growth, namely the dimension of the critical nucleus,
$i$.\cite{Bartelt1992,Family1984,Family1995} This is done by comparing the
experimental size distribution of QDs and its evolution during the first stage
of film formation to the theoretical function that, in turn, depends upon $i$.
To be specific, in the framework of dynamic scaling, the size distribution
function of the number density of islands at coverage $\Theta$ is given by%
\begin{equation}
N_{i}(u)=\frac{\Theta}{\langle s\rangle^{2}}f_{i}\left(  u\right)  \text{,}
\label{pippo}%
\end{equation}
where $u=s/\langle s\rangle$, $\langle s\rangle$ being the island average size
and $f_{i}$ is the scaling function that, according to Amar and Family, reads,
for $i\geq1$: $f_{i}(u)=C_{i}u^{i}e^{-ia_{i}u^{1/a_{i}}}$, where $C_{i}$ and
$a_{i}$ are constants.\cite{Family1995} Dynamic scaling was first introduced
to describe\ 2D islanding and was substantiated by computer
simulations\cite{Bartelt1992,Family1984,Family1995} and experimental studies
both in homo\cite{Stroscio1994,Joyce1997} and
heteroepitaxial\cite{Petroff1995,Jones2002a,Jones2000} growth. Ebiko
\textit{et al.}\cite{Ebiko1998} were the first to show the applicability of
Eq. (\ref{pippo}) to 3D growth in semiconductor heterostructures, provided
that $s$ was interpreted as the volume of the islands. Fig. 2 shows the scaled
island volume distributions for the InAs deposits around the transition, from
$1.45$ to $1.82$ ML. Both families have been included in the data analysis,
taking care to separate the distributions related to the small and large QDs
that are reported in Fig.2(a) and Fig.2(b) and (c) respectively. The scaling
function for small QDs closely resembles that expected for a system with
critical nucleus $i=0$.\cite{Family1995} This implies the adatoms freeze
spontaneously on the surface, giving rise to a nucleation center. Such a
behavior can be explained by the presence of defects on the surface, such as
steps: actually, small QDs nucleate almost exclusively at the upper edge of
steps and 2D islands (Fig.1). This fact indicates the presence of a minimum in
the potential in the proximity of the step edge that makes the monomer stable.
The scaling function does not change between $1.45$ and $1.61$ ML; for higher
coverages the small QD density is negligible in respect to that of large QDs.

As far as large QDs are concerned, we distinguish two different types of
behavior. Up to $1.57$ ML of InAs deposit we observe a kind of mixed
distribution function with an undefined value of $i$, \textit{i.e.} a
transition region where large QDs belonging to $i=2$ (Fig. 2(c)) and $i=0$
coexist. The latter ought to be nothing but the grown small QDs. Starting from
$1.59$ ML (Fig.2(c)) the size distributions change completely, approaching a
shape compatible with $i=2$.\cite{nota3} A new aggregation mechanism is now
operating whereby three atoms are required to form a stable nucleus. Fig.2(d)
displays the comparison between the $f_{2}(u)$ ($C_{2}=1.97$, $a_{2}=0.30$)
and the averaged experimental data; the agreement between the two curves is
excellent. The different aggregation mechanisms allow us to maintain that
large QDs are not merely the direct evolution of the small QDs.

Apparently, the number of atoms required to form a stable nucleus when $i=2$
is three times greater than that for $i=0$. On the basis of this obvious
argument, we must expect that, at the onset of the $i=2$ nucleation mechanism,
an extra amount of free matter (diffusing monomers) is available on the
surface. In Fig.3(a) the total volume of the large QDs is plotted as a
function of InAs coverage. It is highly evident that the increase of the large
dot volume in the transition region implies a quantity of matter well beyond
that provided by the impinging flux $F_{o}$ (lower line in the Fig. 3(a)). To
be precise, in the range $1.6-1.8$ ML the \textit{effective} flux is
$F=4.6F_{o}$. Above $1.8$ ML, the volume increase reverts to being compatible
with $F_{o}$. The extra quantity of matter amounts to roughly $0.9$ ML (Fig.
3(a)). We have already reported\cite{Placidi2005} evidence of step erosion
from QDs nucleated at step edges, setting a lower limit to $0.3$ ML. A recent
work\cite{Jacobi2005a} confirms our finding: by looking at those data, an
amount greater than $0.3$ ML of eroded steps might be estimated. Even though
the erosion could be responsible for the whole supplementary $0.9$ ML, a
further contribution could arise from substrate intermixing and In
segregation.\cite{Patella2003,Patella2004}

The total volume contained in large QDs is determined by the equation:
$V_{large}^{T}=\rho_{large}\langle V_{large}\rangle$, where $\rho_{large}$ is
the density of the large islands and $\langle V_{large}\rangle$ is the mean
volume of the single large island. To specify how the variation of the large
QD density and mean volume contribute to the volume increase of $V_{large}%
^{T}$, we plot separately in Fig. 3(b) the two terms $\frac{d\rho_{large}%
}{d\Theta}\langle V_{large}\rangle$ and $\rho_{large}\frac{d\langle
V_{large}\rangle}{d\Theta}$ as a function of $\Theta$. The derivatives of
$\rho_{large}$ and $\langle V_{large}\rangle$ have been calculated numerically
by interpolating the experimental data. During the first stage of transition
the volume increase is mainly due to the sudden nucleation of large islands
and it is only subsequently that single island growth prevails. At the
transition the QD density explosion is then bound up with the substantial
increase in the adatom density between $1.6$ and $1.8$ ML, as an increase in
the effective deposition flux.

The growth instability leading to the 2D-3D transition is thermodynamic in
character, this being caused, as pointed out by Cullis \textit{et
al.},\cite{Cullis2001,Cullis2002,Cullis2005} by the strain energy relaxation.
The nucleation process begins at $1.45$ ML of InAs deposition and, more
importantly, a single monomer is enough to give rise to a nucleation center.
In accordance with the model proposed by Dehaese \textit{et al.}%
,\cite{Dehaese1995} at $1.45$ ML the average concentration of In in the
uppermost layer is certainly greater than $82\%$.\cite{Patella2004} However,
our data clearly show that the nucleation is preferential at the step edges,
an occurrence which we highlight further. Although the system, from the
thermodynamic point of view, prefers to grow by forming 3D islands, the
conditions for this to occur are met at step edges. On the other hand, at
$1.59$ ML, when the scaling analysis reveals that a stable nucleus needs three
monomers ($i=2$), large QDs first appear at steps and then all over the
surface. Concomitantly, a great amount of monomers becomes available at the
surface because of the step erosion due to small ($i=0$) QDs, which in the
mean time have increased in size. In the framework of the rate equation
approach,\cite{Venables1973,Tomellini1998} the high monomer concentration
$n_{1}$ promotes the nucleation process that is proportional to $n_{1}n_{2}$
($n_{2}$ is the dimer concentration). However, this term competes with the
dimer dissociation term proportional to $n_{2}$. Therefore kinetics implies
that the observed explosive nucleation process can take place if the
production of a high concentration of monomers occurs together with the dimer
dissociation constant much lower than the nucleation constant. The conclusions
are thus apparent: \textit{in order for 2D-3D sudden transition to occur, it
is necessary that both energetic (In concentration at the growing layer) and
kinetic conditions are favorable}.

In summary, we have highlighted the fact that only an appropriate combination
of thermodynamics and kinetics allows the InAs/GaAs(001) heterostructure to
undergo a sudden nucleation and growth of large QDs. Once the appropriate In
concentration has been reached, the $i=0$ nucleation is favored thanks to
steps. Moreover, the subsequent step erosion produces a great amount of fresh
monomers which, in turn, increase the probability of having the $i=2$ (or,
more generally, $i\neq0$) nucleation over the entire surface.

The present work has been partly supported by the FIRB project code No. RBNE01FSWY\_007.

\newpage

\subsection{Figure caption}

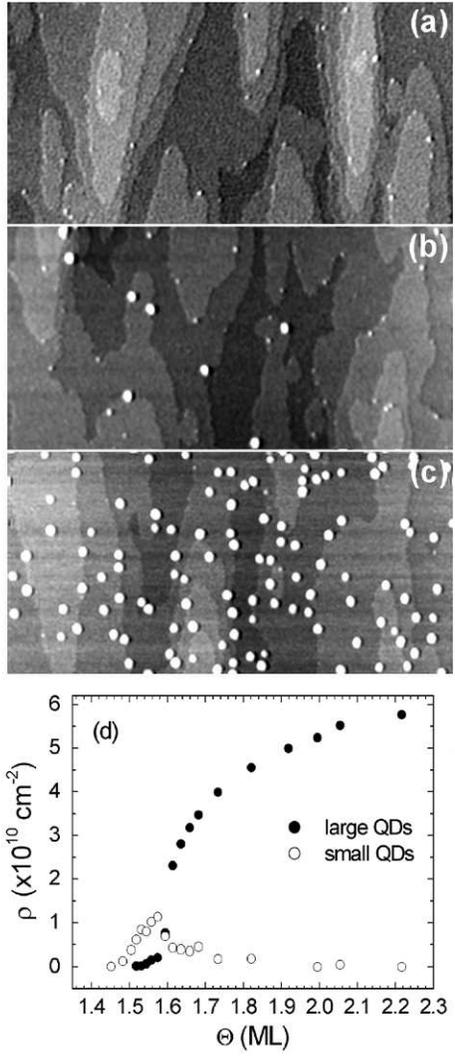
\textbf{Fig. 1} - AFM\ topographies ($1.0$ $%
\operatorname{\mu m}%
$ $\times$ $0.5$ $%
\operatorname{\mu m}%
$) for: $1.54$ ML (a), $1.57$ ML (b), and $1.64$ ML (c) of InAs coverage.
Panel (d)\ shows the number density dependence on InAs coverage of small and
large QDs

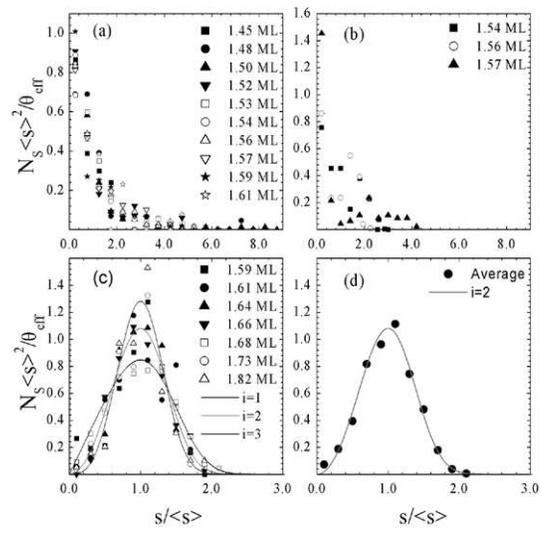
\textbf{Fig. 2} - Scaled distributions of the experimental island volume for:
small QDs (a), large QDs in the range $1.54-1.57$ ML of InAs coverages (b),
large QDs in the range $1.59-1.82$ ML of InAs coverages (c). Solid lines in
panel (c)\ show the theoretical scaling function for $i=1,2,3$. Panel (d)
shows the average of the experimental distributions of panel (c) compared\ to
the theoretical scaling function for $i=2$.

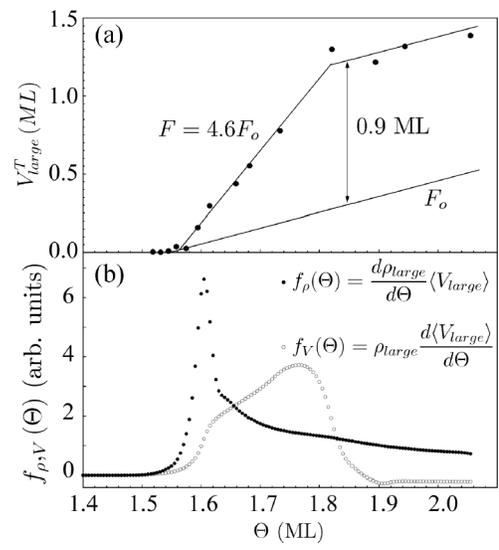
\textbf{Fig. 3} - (a) Total volume $V_{large}^{T}$ of large QDs plotted as a
function of InAs coverage. The lowest line indicates the InAs flux ($F_{o}$)
above the 2D-3D transition. The volume increase in the range between $1.6-1.8$
ML is accounted for by the effective flux $F$ (b) Derivative terms
$\frac{d\rho_{large}}{d\Theta}\langle V_{large}\rangle$ and $\rho_{large}%
\frac{d\langle V_{large}\rangle}{d\Theta}$ of $V_{large}^{T}$ plotted as a
function of inAs coverage.


\begin{thebibliography}{99}
\bibitem {nota1}The sample was placed in a region where the molecular flux
changes linearly with sample position.
\bibitem {nota2}The two 3D structures have been distinguished on the basis of
their $A/V$ ratio, where $A$ and $V$ are the basal area and the volume of the
single island, respectively. As matter of fact, the statistical distributions
of $A/V$ ratio of the islands are bimodal at any InAs coverage.
\end{thebibliography}
\end{document}